\documentclass[conference]{IEEEtran}

\usepackage{cite}
\usepackage{graphicx}
\usepackage{amsmath}
\usepackage{epstopdf}
\usepackage{amsfonts}
\usepackage{amssymb}
\usepackage{subfig}
\usepackage{psfrag}
\usepackage{color}

\begin{document}

\title{Impact of Timing and Frequency Offsets on Multicarrier Waveform Candidates for 5G}

\author{\normalsize Amir Aminjavaheri$^\dagger$, Arman Farhang$^*$, Ahmad RezazadehReyhani$^\dagger$ and Behrouz Farhang-Boroujeny$^\dagger$  
\\$^*$CTVR / The Telecommunications Research Centre, Trinity College Dublin, Ireland, \\
$^\dagger$ECE Department, University of Utah, USA. \\
Email: \{aminjav, rezazade, farhang\}@ece.utah.edu, farhanga@tcd.ie  \vspace{-0.1in} }

\maketitle

\begin{abstract}
This paper presents a study of the candidate waveforms for 5G when they are subject to timing and carrier frequency offset. These waveforms are: orthogonal frequency division multiplexing (OFDM), generalized frequency division multiplexing (GFDM), universal filtered multicarrier (UFMC), circular filter bank multicarrier (C-FBMC), and linear filter bank multicarrier (FBMC). We are particularly interested in multiple access interference (MAI) when a number of users transmit their signals to a base station in an asynchronous or a quasi-synchronous manner. We identify the source of MAI in these waveforms and present some numerical analysis that confirm our findings. The goal of this study is to answer the following question, ``Which one of the 5G candidate waveforms has more relaxed synchronization requirements?''.
\end{abstract}

\section{Introduction}\label{sec:Intro}

In the recent years, multicarrier techniques have been among the most popular and accepted technologies in the wireless broadband communications. Among various multicarrier designs, orthogonal frequency division multiplexing (OFDM) has gained a special attention due to its simplicity and effective equalization structure. In fact, the long-term evolution (LTE) standard, which has OFDM as its underlying physical layer (PHY) modulation technology, offers great data rates and capacities, specially in downlink. However, the shortcomings of OFDM such as its high spectral leakage and strict synchronization requirements, has recently motivated research and industrial communities to propose new waveforms to keep the advantages of OFDM while addressing its drawbacks.

As mentioned above, although LTE offers great opportunities in the downlink, it has some challenges in the uplink due to fundamental drawbacks of OFDM. Specifically, the orthogonality of OFDM is based on strict synchronization between the users, and as soon as the synchronization is lost (e.g., due to multiple access, multi-cell operation or Doppler effects) multiple access interference prevails \cite{wunder20145gnow}. Accordingly, the uplink of LTE is based on resource demanding closed-loop procedures to establish the required synchronization. Furthermore, some of the downlink features such as carrier aggregation are also very limited in the uplink due to the interference issues \cite{iwamura2010carrier}. The aforementioned challenges become more serious when considering the technological requirements of the fifth generation (5G) of cellular networks \cite{andrews2014will}. In fact, some emerging applications in 5G such as the smart city and Internet of Things (IoT), by definition, need to support many machine-type communication (MTC) nodes with the design criteria of low implementation cost, long battery life, and extremely low latency message delivery \cite{wunder20135gnow}. This drives the idea of \textit{asynchronous communication} in order to avoid the problems of LTE such as its high latency due to the network overhead imposed by the sophisticated training signaling schemes \cite{wunder20145gnow, berg2014multiuser, schaich2014relaxed}. 

We also note that OFDM waveform is built based on a rectangular pulse shape/prototype filter,  and uses a cyclic prefix (CP) to simplify the equalization. Although the rectangular pulse shape is well-localized in time, it is poorly localized in frequency due to the abrupt transitions in symbol boundaries. This is the major source of OFDM sensitivity to synchronization errors in the uplink. Solutions to moderate this problem of OFDM by application of widows at both transmitter and receiver sides have been well studied, e.g., \cite{farhang2008multicarrier, muller2001optimum}.  The first contribution of this paper is to show that the OFDM deficiencies arising from the use of rectangular windows extend to the new waveforms that are reviewed below (although to a lesser extent), and hence the use of windowing methods in these new waveforms builds on the same concept. 

Filter bank multicarrier (FBMC) is another candidate that can achieve time-frequency localization by utilizing a well-designed pulse shape/prototype filter \cite{farhang2014filter, farhang2011ofdm}. Furthermore, to maintain the orthogonality in such systems, real and imaginary symbols are staggered in time and frequency \cite{sahin2014survey}. Despite good time-frequency localization, FBMC has its own drawbacks. Specifically, application of FBMC to multiple-input multiple-output (MIMO) channels is limited \cite{farhang2011ofdm} and also the ramp-up and ramp-down of the FBMC signal at the beginning and the end of each packet reduces its bandwidth efficiency in applications that demand communication of short bursts, e.g., in MTC. 

In order to overcome the above problems of FBMC, circularly pulse shaped waveforms such as generalized frequency division multiplexing (GFDM) and circular FBMC (C-FBMC) have emerged \cite{michailow2014generalized, lin2014advanced}. In GFDM, complex QAM symbols are modulated using time and frequency localized pulses based on the Gabor system \cite{matthe2014generalized}. However, as a consequence of the Balian-Low theorem \cite{daubechies1990wavelet}, orthogonality cannot be achieved in GFDM, which makes GFDM a non-orthogonal waveform. C-FBMC combines the ideas of real/imaginary staggering and circular pulse shaping in order to maintain the orthogonality as well as all the advantages of GFDM. Although circularly pulse shaped waveforms enhance the bandwidth efficiency of FBMC, they cannot achieve the same insensitivity to synchronization errors compared to the linear FBMC. The second contribution of this paper is to show this fact.

Another candidate waveform for 5G that has been recently proposed is the universal filtered multicarrier (UFMC) \cite{vakilian2013universal, schaich2014relaxed}. UFMC modifies OFDM by applying a filter on each group of subcarriers, i.e., a physical resource block, in the context of LTE systems. This improves the robustness to synchronization errors by limiting the out of band emissions of the subcarriers.

The impact of multiuser synchronization errors including symbol timing offset (TO) and carrier frequency offset (CFO) on the performance of OFDM and FBMC systems has been studied extensively in the past, e.g., \cite{park2003performance, raghunath2009sir, hashemizadeh2014sensitivity,mattera2015analysis, fusco2008sensitivity, saeedi2010sensitivity}. Moreover, a performance comparison between UFMC and OFDM with respect to TO and CFO has been recently presented in \cite{schaich2014relaxed, vakilian2013universal}. However, there exists no study in the literature investigating the TO and CFO effects on all the major 5G candidate waveforms. In this paper, we scrutinize the performance of all the proposed candidate waveforms for 5G in presence of timing and frequency offsets in an attempt to answer the question 
``Which one of the 5G candidate waveforms has more relaxed synchronization requirements?''. Accordingly, the aim in this paper is to (i) provide a clear-cut explanation on the behavior of the aforementioned 5G contender waveforms with respect to multiuser timing and frequency misalignments; (ii) compare their robustness with each other; and (iii) highlight the fact that the  linear FBMC has the least sensitivity to TO and CFO.

The rest of this paper is organized as follows. In Section \ref{sec:System_Model}, we explain the multi-user system model in uplink of a cellular communication system. In Section \ref{sec:TO_Analysis}, the phenomena of spectral leakage due to timing misalignment is explained and the behavior of different multicarrier waveforms with respect to multiuser TO is studied. Section \ref{sec:CFO_Sensitivity} investigates the impact of multiuser frequency misalignment on multicarrier waveforms and compares their performance with each other. Section \ref{sec:BER} provides an evaluation on the performance of different waveforms when a combined effect of CFO and TO exists. Concluding remarks are drawn in Section \ref{sec:Conclusion}.

\section{Uplink System Model} \label{sec:System_Model}
Any multicarrier scheme can be thought of as a mapping between the message space and the signal space using a basis that spans in both the time and frequency domain \cite{sahin2014survey}. Moreover, one can successfully recover the message at the receiver, if the mapping is one-to-one. Here the scenario that we are considering is the uplink transmission direction of a multicarrier frequency division multiple access (FDMA) system, where the bandwidth is partitioned into the total of $N$ subcarriers and each user accesses a cluster of subcarriers. Thus, for the $\ell^\mathrm{th}$ user, the transmit symbol $X_{mk}^{(\ell)}$ corresponds to the data symbol $D_{mk}^{(\ell)}$ according to 
\begin{eqnarray}
X_{mk}^{(\ell)} = 
\begin{cases}
D_{mk}^{(\ell)}, & k \in \mathcal{N}_\ell\\
0,				& k \notin \mathcal{N}_\ell
\end{cases}
\end{eqnarray}
where $k$ is the subcarrier index, $m$ is the symbol time index, and $\mathcal{N}_\ell$ denotes the set of subcarrier indices that are devoted to the $\ell^\mathrm{th}$ user. Consequently, we can represent the transmit signal of the $\ell^\mathrm{th}$ user by
\begin{eqnarray}
x_\ell(t) = \sum_{m=-\infty}^{+\infty} \sum_{k=0}^{N-1} X_{mk}^{(\ell)} g_{mk}(t)
\end{eqnarray}
where $g_{mk}(t)$ is the transmitter basis function corresponding to the $(m,k)$ time-frequency point. 

After propagating through the channel, and assuming a TO error of $\tau_\ell$ and a CFO error of $\varepsilon_\ell$ between the $\ell^\mathrm{th}$ transmitter and the base station (BS), the received signal at the BS can be expressed as 
\begin{eqnarray}
y(t) = \sum_{\ell} y_{\ell}(t - \tau_\ell) e^{j 2 \pi \varepsilon_\ell t/T} + n(t)
\end{eqnarray}
where 
\begin{eqnarray}
y_\ell(t) = \int_\tau c_\ell (\tau, t) x_\ell (t - \tau) d\tau,
\end{eqnarray}
is the signal of the $\ell^\mathrm{th}$ user distorted by the multipath time-varying channel impulse response $c_\ell(\tau, t)$, $n(t)$ is the AWGN, and $T$ denotes the symbol period.

In this paper, in order to focus on the effects of timing and frequency misalignments of FDMA users, we ignore the fading effect of the channel and assume ideal channel response. Moreover, given the $\ell^\mathrm{th}$ user is the user of interest, we assume $\tau_\ell=0$, and $\varepsilon_\ell = 0$, meaning the BS is synchronized with the signal coming from the user of interest. However, signals of other users are not perfectly aligned with respect to the receiver. Based on this plot, the transmitted data symbols of the user of interest can be recovered according to
\begin{IEEEeqnarray}{rCl} 
\hat{X}_{mk}^{(\ell)} &=& \langle y(t), h_{mk}(t) \rangle \label{eq:rcv_inner_product}\\
&=& X_{mk}^{(\ell)} + I_\mathrm{MAI} + \eta
\end{IEEEeqnarray}
where $h_{mk}(t)$ represents the receiver basis corresponding to the $(m,k)$ time-frequency point. If the underlying modulation scheme is orthogonal, e.g. OFDM, FBMC, etc., $h_{mk}(t) = g_{mk}(t)$. However, for a non-orthogonal waveform, e.g. GFDM, the receiver basis may not be the same as the transmitter basis in general. Time and frequency misaligned symbols of other users will cause a multiple access interference (MAI) which we have denoted by $I_\mathrm{MAI}$. The noise contribution is also denoted by $\eta$.

\section{TO Sensitivity Analysis}\label{sec:TO_Analysis}
This section focuses on the timing offset sensitivity analysis of different candidate waveforms for the physical layer of 5G systems. To pave the way for a better understanding of our analysis in Section~\ref{subsec:TO_Analysis}, some general concepts on timing offset effects are first discussed in Section~\ref{subsec:spectral_leakage}.

\subsection{Spectral Leakage Due to Timing Offset} \label{subsec:spectral_leakage}
In any multicarrier waveform, as far as a particular symbol time index is concerned, an observation window with a finite duration is applied to the received signal and the processing is performed on that finite interval. This fact can be realized from (\ref{eq:rcv_inner_product}) and it can be noticed that the receive pulse shape (windowing function), $h_{mk}(t)$, has a finite duration in practice. This finite windowing leads to some peculiarities when a symbol timing misalignment exists between the frequency separated users. The objective of this section is to explain this phenomenon, hence, facilitate understanding of the behavior of different waveforms and, accordingly, develop the respective sensitivity  in Section~\ref{subsec:TO_Analysis}. 

We use $T$ to denote the symbol period, and $T_W$ for the duration of the window for data detection at the receiver. We also note that it is common to choose $T_W = KT$, where $K$ is an integer. In filter bank literature, $K$ is called the overlapping factor. For OFDM $K=1$, for UFMC $K=2$ (including the padded zero for analysis), and for filter bank based waveforms $K$ is a design factor that indicates the number of overlapping symbols in the time domain. Since the processing is limited to the signals with the duration of $T_W$, Fourier series analysis can be employed in order to explain the behavior of different waveforms with respect to the timing misalignment. Based on the Fourier series analysis, an intuitive approach to understand (\ref{eq:rcv_inner_product}) is to think of the windowed signal as a \textit{single period} of a periodic signal. In (\ref{eq:rcv_inner_product}), if perfect reconstruction is assumed when the users are fully synchronous, $X_{mk}^{(\ell)}$ can be retrieved free of MAI. However, in presence of timing offset, the periodic extension of the signals from other users exhibits discontinuities which cause non-zero projections on the frequencies belonging to the user of interest.

\begin{figure}[!t]
\centering
\includegraphics[scale=0.6]{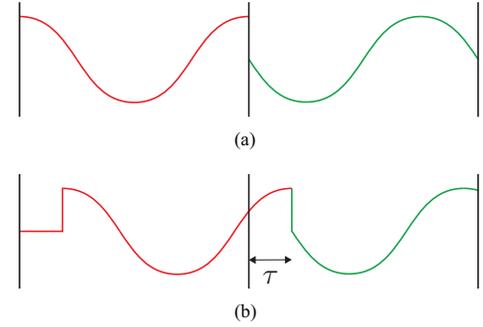}
\caption{Illustration of signal discontinuities due to TO.}
\vspace{-0.1in}
\label{fig:Discontinuities}
\end{figure}

To illustrate the above point, consider Fig. \ref{fig:Discontinuities}a. This  represents two sinusoids that are perfectly time aligned with their corresponding observation windows. In this case, the periodic extension of each observation window is also a pure sinusoid and, thus, it exhibits a non-zero projection onto the respective basis function/frequency and results in a zero projection onto the other basis functions/frequencies. Fig. \ref{fig:Discontinuities}b, on the other hand, illustrates the case of having a TO error, $\tau$, between the observation window and the received signal. In this case, the periodic extension of each observation window exhibits discontinuities at three different points; Two at the boundaries and one at the intersection of two symbols. These discontinuities cause non-zero projections on the entire Fourier series expansion, and thus, \textit{spectral leakage} due to symbol timing misalignment occurs. This spectral leakage is the source of interference between the FDMA users in the uplink of a cellular system that are asynchronous in time.

To decrease the level of discontinuities and hence reduce the impact of spectral leakage, windows with smooth tails should be utilized at both the transmitter and the receiver. When such a window is used in the transmitter, at the receiver side, the discontinuities at the symbol intersection points disappear. On the other hand, applying a window at the receiver will remove the  discontinuities at the window boundaries. 

\subsection{Sensitivity of Different Waveforms to Timing Offset}\label{subsec:TO_Analysis}
As discussed above, the amount of multiuser spectral leakage due to TO for any multicarrier waveform is closely related to the discontinuities in the observation window at the receiver. Moreover, filtering at both transmitter and receiver sides helps to mitigate the performance loss caused by the timing offset. Based on this discussion, it is clear that since UFMC adds a filtering operation to OFDM at the transmitter, it has a better ability to resolve the multi-user TO interference than OFDM. Moreover, FBMC can perform even better, thanks to its smooth filters both at the transmitter and receiver. 

\begin{figure}[!t]
\centering
\includegraphics[scale=0.65]{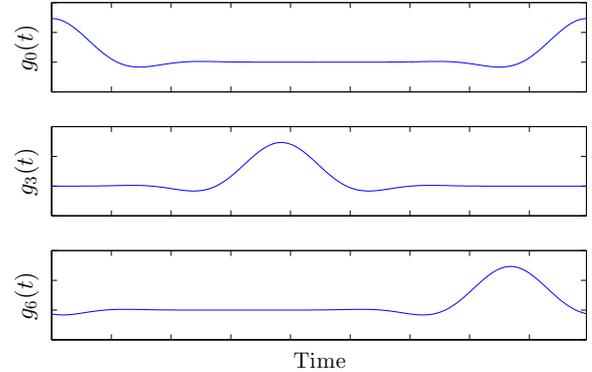}
\caption{Circular filters in GFDM.}
\vspace{-0.2in}
\label{fig:circular_filters}
\end{figure}

In waveforms with linear pulse shaping, such as FBMC and UFMC, the same pulse shape is used for symbols at different time indices, and thus there is a symmetry in the TO performance of different symbols of a packet. Interestingly, this is not the case for circularly pulse shaped waveforms. To better understand this phenomenon, consider Fig. \ref{fig:circular_filters}, where we have depicted typical pulse shapes of the $1^{\mathrm{st}}$, $4^{\mathrm{th}}$ and $7^{\mathrm{th}}$ symbol indices of a GFDM packet consisting a total of seven symbol periods. Here, although $g_{3}(t)$, and in general, the central symbols can effectively bring the boundary discontinuities close to zero, the symbols located at the edges of the packet, specially $g_{0}(t)$, cannot achieve the boundary continuity in presence of timing offset. This results in higher performance degradation in the symbols located at the edges of the data packet as compared to the central ones.

Besides the boundary discontinuities, another discontinuity also occurs inside the receive window, as discussed earlier. Since other pulses than $g_0(t)$ have zero-merging tails, the magnitude of this discontinuity is closely related to $g_0(t)$, which has large magnitudes at the edges of the packet. Accordingly, the cause of this discontinuity is mainly due to the symbol corresponding to $g_0(t)$ being shifted as a result of the timing misalignment. In this case, at the receiver, the symbol that its main lobe coincides with the position of the discontinuity undergoes the maximum spectral leakage. Other symbol indices bring that discontinuity close to zero and thus may not experience spectral leakage to that extent. This suggests that to reduce the amount of MAI induced by TO in GFDM and C-FBMC, one remedy could be to turn off the first symbol of the packet. Obviously, this leads to some loss in spectral efficiency. This method has been applied in \cite{matthe2014influence} in the context of reducing out-of-band (OOB) emission. Here, we emphasize that this also has the impact of reducing MAI at the receiver.   

Next, we present some numerical results that confirm the above observations and provide more insight to the sensitivity of different waveform to TO. We consider an uplink scenario with two users. The total number of subcarriers for all waveforms is 256. From these a total of 72 subcarriers are used; 36 contiguous subcarriers are allocated to each user. There is a guard subcarrier between the two users' subcarriers.  Perfect power control is assumed for the users. For OFDM, C-FBMC, and GFDM, a CP length of $32$ samples is used. Seven symbols are transmitted in each GFDM or C-FBMC packet. In UFMC, a Dolph-Chebyshev filter with the length $33$ and stop band attenuation of 40 dB is used. In the both cases of FBMC and C-FBMC the Mirabbasi-Martin filter (a.k.a the PHYDYAS filter) \cite{mirabbasi2003overlapped, bellanger2010fbmc} is used. For GFDM, we have used  a root raised-cosine filter with the roll-off factor of $\alpha=0.4$. We only consider an AWGN channel, hence, channel does not introduce any time spreading and as a result the whole range of CP introduces a time period where OFDM, GFDM, and C-FBMC remain insensitive to TO. As the following numerical results show, this statement is not quite an exact one for GFDM. This is a result of the fact that GFDM is a non-orthogonal waveform and as a result the ZF detector (that we use here) has some peculiar behaviors whose details are beyond the scope of this paper and remains as a future study.

One of the two users is considered as the user of interest, and it is assumed that the BS is time-aligned with the signal coming from this user. A TO is applied to the second user and the impact of this TO on the MAI seen by the first user is studied.  In Fig.~\ref{fig:sim_MAI_2}, we have compared the MAI power, $P_{\mathrm{MAI}} = E \{| I_\mathrm{MAI} | ^ 2\}$, for different values of TO in OFDM, UFMC, FBMC, C-FBMC, and GFDM waveforms.

\begin{figure}[!t]
\centering
\includegraphics[scale=0.55]{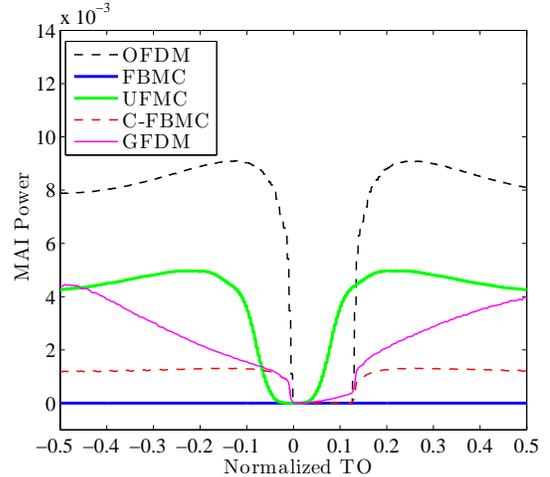}
\caption{MAI power as a function of TO for different waveforms.}
\vspace{-0.2in}
\label{fig:sim_MAI_1}
\end{figure}

Considering the results in Fig.~\ref{fig:sim_MAI_2}, the following observations are made. (i) OFDM is the most sensitive waveform to TO. (ii) FBMC has almost no sensitivity to TO, as noted before, thanks to combined filtering at both the transmitter and receiver sides. (iii) The CP interval introduces a time zone that can absorb TO errors. As noted above, when TO is within this range, OFDM, GFDM, and C-FBMC remain insensitive to TO. Presence of time spreading in the channel, clearly, reduces this range, (iv) UFMC also exhibits a TO insensitive range. This corresponds to small tails of the transmitter filter in UFMC.

\section{CFO Sensitivity Analysis}\label{sec:CFO_Sensitivity}
OFDM systems, in uplink, are known to be highly sensitive to CFOs between different users. Therefore, new waveforms proposed for 5G systems have to provide a much lower sensitivity to CFOs compared with OFDM systems. Imperfect frequency domain synchronization among different uplink users leads to MAI caused by inter-carrier interference (ICI). To reduce the CFO induced MAI and hence provide a more relaxed synchronization requirements than OFDM, the candidate waveforms for 5G strive to localize their subcarriers in frequency. This directly impacts the leakage from asynchronous subcarriers to the others.

In this study, we consider block subcarrier allocation scheme where a block of contiguous subcarriers is allocated to each user and one guard subcarrier is considered between different users. Due to the linear pulse shaping, each user's block in FBMC and UFMC is highly localized in the frequency domain, given that a well designed prototype filter is deployed. Thus, the amount of leakage among different users due to the CFOs is negligible. Based on the results of \cite{FarhangICC15}, due to the circular pulse shaping, GFDM and C-FBMC signals can be thought as superposition of a number of tones that are scaled by the data symbols as well as frequency response of the prototype filter. More specifically, each data bearing subcarrier in such systems include $2K-1$ frequency samples scaled with prototype filter coefficients in the frequency domain where $K$ is the number of symbols in each data packet. Due to the presence of the rectangular window in such systems, each frequency sample, i.e., a tone, can be represented by a sinc function in frequency domain. It is worth mentioning that all the zero crossings of the sinc functions due to different subcarriers in a fully synchronous system are coinciding. If the subcarrier spacing is $\Delta f$, the spacing between different tones in such systems is $\frac{\Delta f}{K}$. Consequently, a normalized CFO of $\varepsilon$ circularly rotates the sinc functions by the relative CFO of $K\varepsilon$ in the frequency domain. 

\begin{figure}
\centering
\includegraphics[scale=0.55]{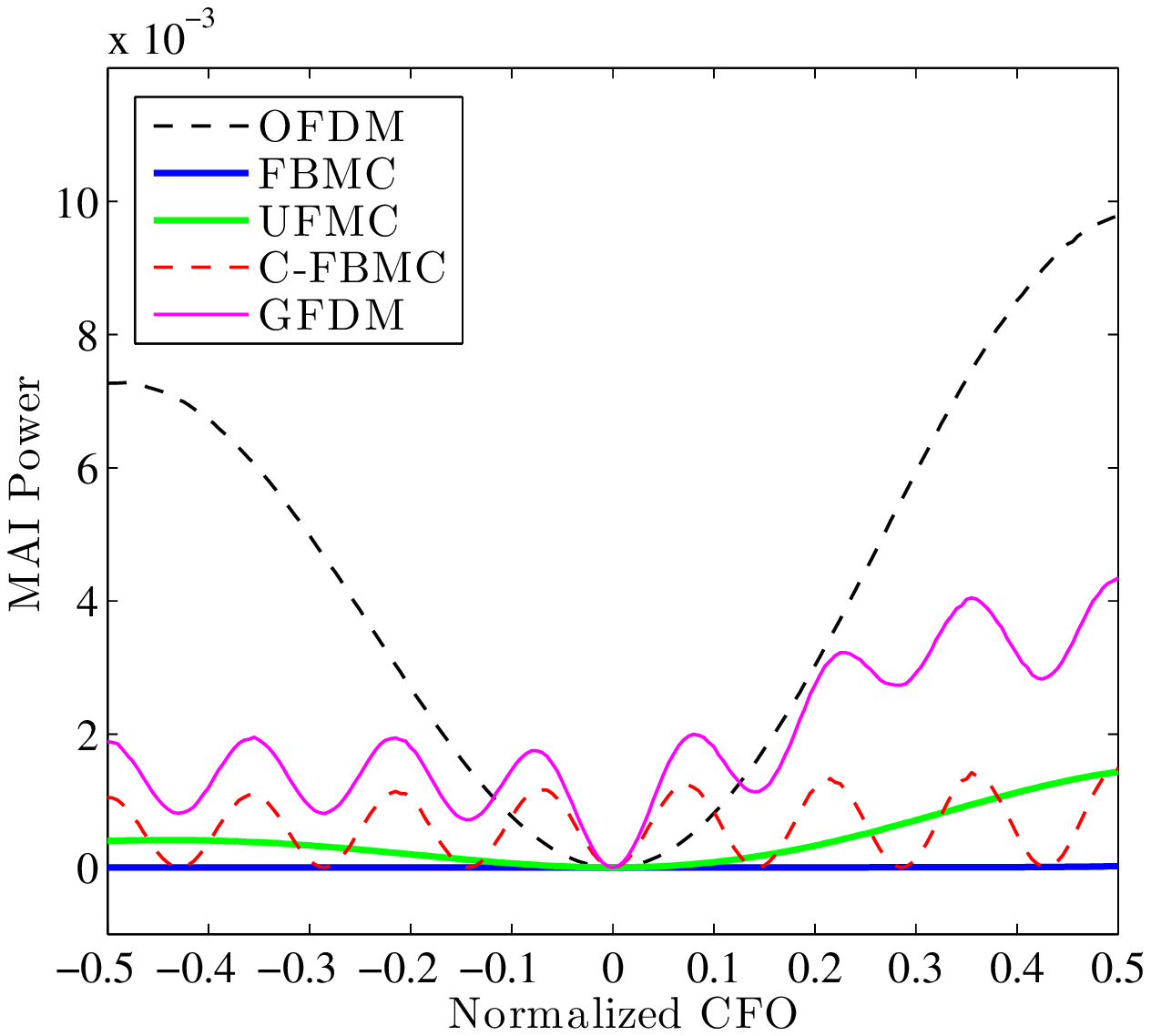}
\caption{MAI power as a function of CFO for different wave- forms.}
\label{fig:sim_MAI_2}
\vspace{-0.2in}
\end{figure}

Fig.~\ref{fig:sim_MAI_2} compares sensitivity of different waveforms to CFO in the uplink of a multiuser system with two active users. The same as in our TO sensitivity analysis, total number of $N=256$ subcarriers is considered. Both users occupy the same bandwidth, each consisting of $36$ subcarriers with a guard band subcarrier in between the users. Perfect power control is also assumed. It is worth mentioning that the overlapping factors of $K=4,~7$ and $7$ are considered for FBMC, C-FBMC and GFDM, respectively. To analyze the CFO effect, the user of interest is perfectly synchronized while the other user has the normalized CFO of $\varepsilon$ that varies in the range $\left[-0.5,0.5\right]$. In our analysis, we compare the MAI power of different waveforms with respect to the normalized CFO. Since all the candidate waveforms for 5G are in the quest for a higher robustness against CFOs than OFDM, in multiuser scenarios like uplink communications, we set the MAI curve of OFDM as a reference.

\begin{figure}
\centering
\includegraphics[scale=0.6]{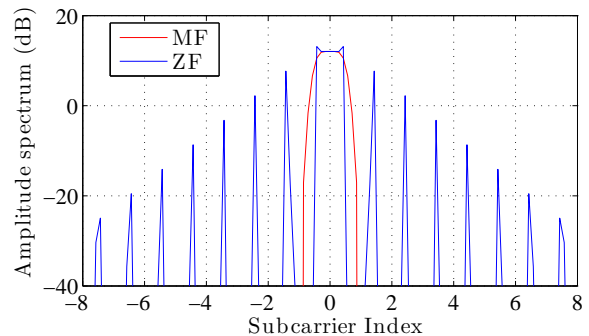}
\caption{Amplitude spectrum of the receiver matched filter in C-FBMC and zero-forcing detector in GFDM.}
\label{fig:MF_ZF_Freq}
\vspace{-0.2in}
\end{figure}

As can be seen from Fig.~\ref{fig:sim_MAI_2}, FBMC and UFMC have a much higher robustness to CFO compared with OFDM. In contrast, GFDM and C-FBMC are more sensitive to CFO than OFDM around the range $\varepsilon\in\left[-0.1,0.1\right]$ and they have a higher robustness to CFOs than OFDM outside that range. One may wonder about the reason for periodic behavior of MAI in C-FBMC and GFDM. As noted earlier, both C-FBMC and GFDM are based on transmission of $2K-1$ tones per subcarrier and overlapping of $K-1$ of them at each side of a given subcarrier. CFO of $\varepsilon$ circularly rotates these tones by $K\varepsilon$ and hence there are points where $K\varepsilon$ is an integer, i.e., $\varepsilon\approx \pm0.143, \pm0.286, \pm0.428$ in Fig.~\ref{fig:sim_MAI_2}. Due to the real orthogonality in C-FBMC signal and the shape of the matched filter that is shown in Fig.~\ref{fig:MF_ZF_Freq}, when $K\varepsilon$ is an integer, zero crossings of the sinc pulses coincide and therefore as long as the frequency samples of different users do not overlap, no MAI is present. This is the reason for the periodic behavior of MAI in C-FBMC. The same as in C-FBMC, the zero crossings of the sinc pulses in GFDM are coinciding when $K\varepsilon$ is an integer. However, the MAI is not zero in this case which is due to the particular shape of the zero forcing (ZF) filter that is used at the receiver side. From Fig.~\ref{fig:MF_ZF_Freq}, one may realize that when $K\varepsilon$ is integer, ZF filter takes samples from wrong discrete Fourier transform (DFT) bins outside its main lobe, which extends up to $8$ subcarriers at each side. This is the source of residual MAI in GFDM for integer values of $K\varepsilon$. Another observation from Fig.~\ref{fig:sim_MAI_2} is that FBMC has the highest CFO robustness compared with other waveforms while GFDM is the most sensitive one after OFDM. UFMC seems to be the second best candidate waveform among the ones analyzed in this paper from CFO robustness point of view.

\section{Putting All Together} \label{sec:BER}

\begin{figure}[!t]
\centering
\includegraphics[scale=0.55]{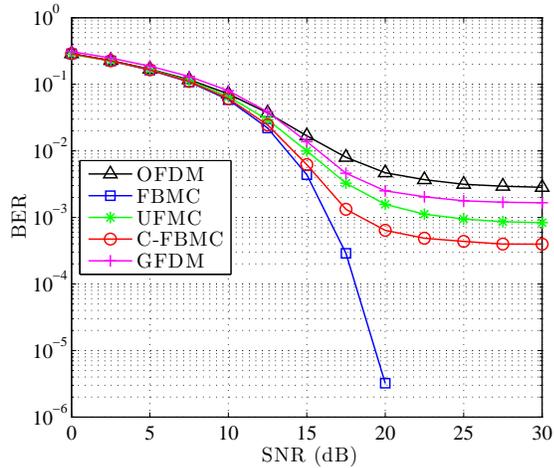}
\caption{BER performance of different waveforms. The normalized TOs and CFOs are selected randomly between $-0.5$ and $+0.5$.}
\vspace{-0.2in}
\label{fig:BER_1}
\end{figure}

\begin{figure}[!t]
\centering
\includegraphics[scale=0.55]{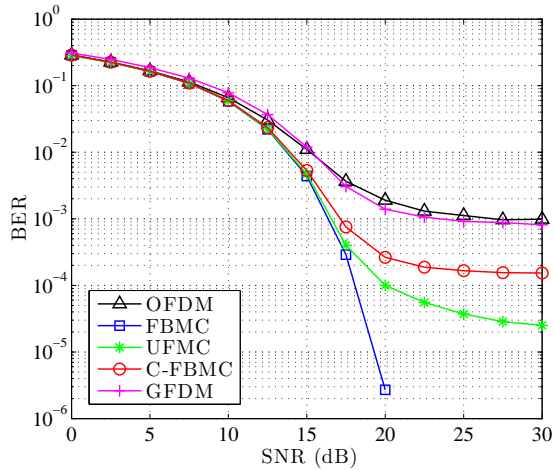}
\caption{BER performance of different waveforms. The users are quasi-synchronous in time but the CFO errors are selected randomly between $-0.5$ and $+0.5$.}
\vspace{-0.2in}
\label{fig:BER_2}
\end{figure}

In this section, we evaluate the combined effect of timing and frequency misalignments for different waveforms using computer simulations. Fig.~\ref{fig:BER_1} presents the uncoded bit error rate (BER) for different waveforms. Here, 5 active users are considered, and the user of interest is assumed to be the middle one in the frequency. The values of normalized TOs and CFOs for other users are selected randomly between $-0.5$ and $+0.5$, which represents a complete asynchronous scenario. Data symbols are from a 16-QAM constellation. All other simulation parameters are the same as before. In Fig.~\ref{fig:BER_2}, we assumed the users are \textit{quasi-synchronized} in time, meaning TOs for CP-based waveforms are in the range of CP, for UFMC are in the range of $\left[ -0.02, +0.02 \right]$, and for FBMC are in the range of $\left[ -0.5, +0.5 \right]$ (i.e., a complete asynchronous scenario). This ensures that the MAI due to timing misalignment is negligible. However, since CFOs are selected randomly between $-0.5$ and $+0.5$, MAI due to frequency misalignments remains.

\section{Conclusion} \label{sec:Conclusion}
We identified the sources of interference when multicarrier systems are subject to TO and CFO. The studied waveforms are OFDM, GFDM, C-FBMC, UFMC, and linear FBMC. It was noted that to reduce sensitivity to TO and CFO, windows with smooth edges should be applied at both the transmitter and receiver sides. Among the above waveforms only linear FBMC satisfies this condition. Second to linear FBMC is UFMC where a window with smooth edges is used at the transmitter. OFDM, GFDM, and C-FBMC fail our tests as in their conventional form, they lack windows with smooth transitions at both the transmitter and receiver sides. However, improvements are possible by taking note of the points discussed in this paper and applying the necessary windows.
\vspace{-0.1in}

\bibliographystyle{IEEEtran}
\bibliography{IEEEabrv,bibFile}

\end{document}